\def\Lsun{$L_\odot$}
\def\Msun{$M_\odot$}
\def\Mv{$M_{\rm v}$}
\def\Teff{$T_{\rm eff}$\ts}
\def\0BMV{$(B~-~V)_{\rm 0}$}
\def\BMV{$B~-~V$}
\def\Z{$Z$}

\def\simgt{\lower.5ex\hbox{$\; \buildrel > \over \sim \;$}}
\def\simlt{\lower.5ex\hbox{$\; \buildrel < \over \sim \;$}}

\documentclass{aa}
\usepackage{graphicx}
\begin{document}
\input psfig.sty
   \thesaurus{05(08.08.2;    
		 08.16.4;    
		 10.07.2;    
		 08.03.2)}   
\title{Mixing, Enhanced Helium and Blue Tails in Globular Clusters}

\author{V. Caloi \inst{1}}

   \offprints{V. Caloi}
   \mail{caloi@saturn.ias.rm.cnr.it}

   \institute{Istituto di Astrofisica Spaziale C.N.R., Via Fosso del
	      Cavaliere, I-00133 Roma, Italy
	     }

   \date{Received ; accepted }

   \titlerunning{Mixing and Blue Tails}
   \authorrunning{V. Caloi }
   \maketitle

   \begin{abstract}
{We investigate the consequences of an increase in the envelope helium
abundance of pre-helium flash red giants in globular clusters, an occurrence
suggested by chemical peculiarities in many red giant atmospheres. Comparing
predictions with the CM diagrams of a few crucial globular clusters, one
finds no evidence for a substantial increase in the surface helium content of
horizontal branch members of these clusters, at least for objects in the RR
Lyrae region or close to it. The possibility that the most peculiar giants
belong to the asymptotic giant branch is discussed. The consequences of a
delay in the helium flash are briefly examined. }

    \end{abstract}


\section{Introduction}
The evolutionary phase so called of the "horizontal branch" (HB) for
Population II stars is well known to depend strongly on the conditions
inherited from the evolution along the giant branch (GB). Phenomena that do
not alter the overall appearance of the GB on the CM diagram
may influence substantially the morphology of the HB. Mass loss is probably
the most important of these events, to which the possibility of noticeable
changes in the chemical composition has recently been added.

In the following we shall consider some of the consequences that deviations
from standard GB evolution may have on HB structures. In particular, we
shall examine the effect of enhanced helium content in HB structures and
of a delay of the helium flash at the GB tip. Both questions are relevant
for the ``second parameter'' problem in globular clusters; the former
has been investigated recently by many authors (see, f.e., the review by
Kraft 1994).

\section{HB models with enhanced helium in the envelope}

The high precision of recent spectroscopic observations of globular cluster
red giant stars allowed to confirm the existence of objects with peculiar
chemical abundances - not expected from current evolutionary models - and to
better define the percentages of peculiar atmospheres and the degree of the
peculiarity for single objects (see the reviews by Smith 1987, Suntzeff
1993 and Kraft 1994). The
phenomenon has been interpreted in terms of "deep" mixing, that is, of the
mixing of envelope material well into the hydrogen burning shell, in regions
where proton captures take place (Denisenkov \& Denisenkova 1990, Langer \&
Hoffman 1995, Cavallo et al. 1996, Pilachowski et al. 1996, Shetrone 1996).

\begin{figure}
\hspace{1cm}
\psfig{figure=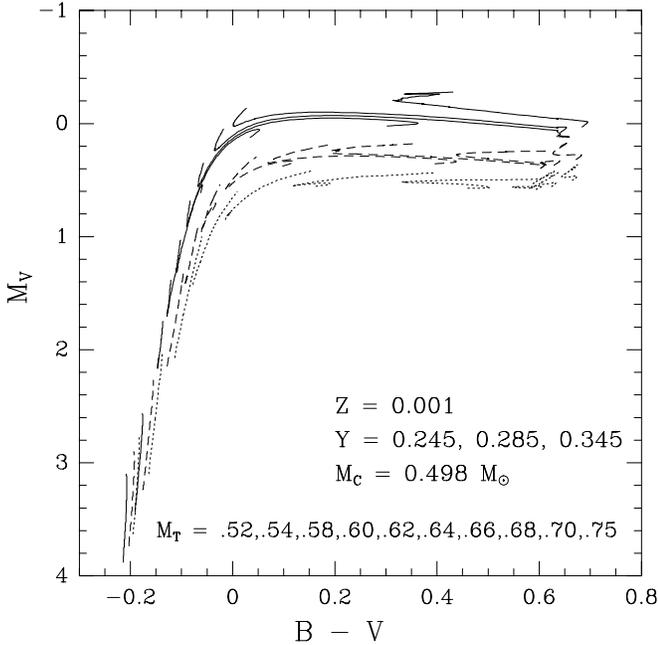,height=9cm}
\caption{HB evolutionary tracks of models with the same core mass and
envelope $Y$ = 0.245 (dotted line), 0.285 (dashed line) and 0.345
(solid line). The tracks end when $Y_{\rm c}$ = 0.1.}
\label{H2445F1}
\end{figure}

One of the consequences of this complex phenomenon should be the increase of
the helium abundance in the envelope of a star that suffers episodes of deep
mixing (Langer \& Hoffman 1995). Such an increase seems assured when Al is
produced at the expense of Mg, but Langer et al. (1997) argue that helium is
enhanced also when O and Ne are substantially depleted.

Sweigart (1997) has examined the impact of helium mixing on the morphology
of the HB. The process of helium enrichment in the envelope
gives rise to an increase in the tip luminosity attained by the red giant
star, and so to an increase in the mass lost during the giant branch ascent.
Besides, an enhanced helium in the envelope means a bluer HB position at a
given total mass (the helium core mass remains almost unchanged).

In Fig. \ref{H2445F1} tracks with $Z$ = 0.001 and three values of the
amount of helium
in the envelope ($Y$ = 0.245, 0.285 and 0.345) are shown (code ATON2.0, see
Ventura et al. 1998; transformations from the theoretical plane through
Kurucz 1993). The tracks illustrate the main HB phase until central helium
$Y_{\rm c}$ = 0.1. A single ``breathing pulse'' (f.e., Castellani et al.
1985) was sometimes encountered when $Y_{\rm c}$ $\sim$ 0.01 (duration
about 1 -- 3 Myr). Since the phenomenon is generally considered spurious
(Dorman \& Rood 1993, Caloi \& Mazzitelli 1993), they have been eliminated
from the tracks shown in the figures.

\begin{figure}
\hspace{1cm}
\psfig{figure=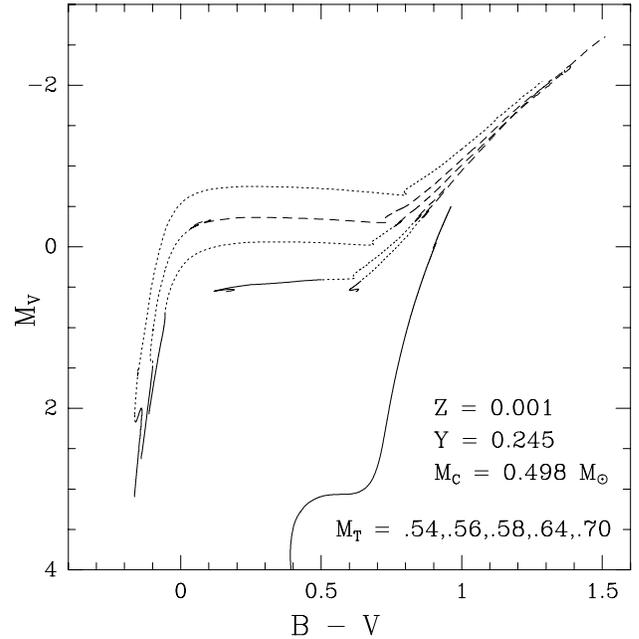,height=9cm}
\caption{Evolution out of the HB and up to the AGB for models with
envelope $Y$ = 0.245; dotted lines indicate fast phases, for $M$ $\geq$ 0.56
\Msun, dashed lines indicate the helium shell burning phase.
The track of a 0.9 \Msun\ red giant is shown as reference.}
\label{H2445F2}
\end{figure}

\begin{figure}
\hspace{1cm}
\psfig{figure=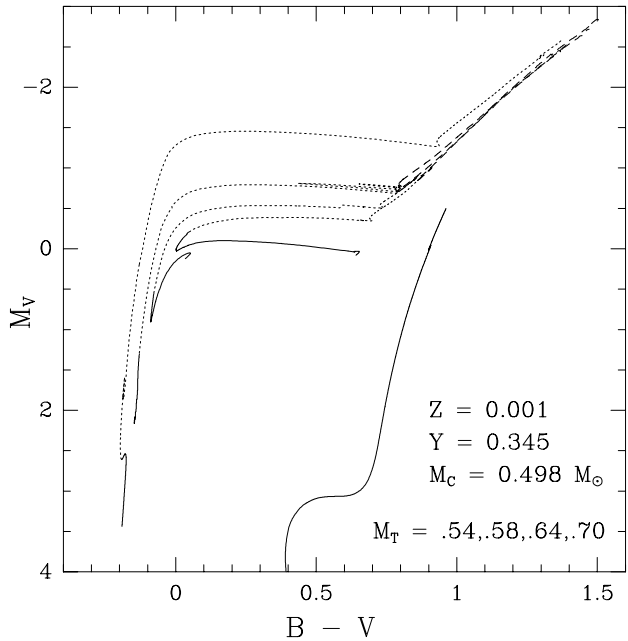,height=9cm}
\caption{Evolution out of the HB and up to the AGB for models with
envelope $Y$ = 0.345; dotted lines indicate fast phases, for $M$ $\geq$ 0.58
\Msun, dashed lines indicate the helium shell burning phase.
The track of a 0.9 \Msun\ red giant is shown as reference.}
\label{H2445F3}
\end{figure}

The case $Y$ = 0.245 is taken to correspond to the normal
helium content, after the envelope increase due to the first dredge-up. In
Figs. \ref{H2445F2} and \ref{H2445F3} the tracks are shown for some masses
up to the asymptotic giant branch (AGB) for the cases $Y$ = 0.245 and
0.345, respectively. The track of a giant of 0.9 \Msun\ ($Z$ = 0.001, $Y$ =
0.235) is added for reference.

The main evolutionary characteristics appear the following.
The mass distribution along the ZAHB does not change substantially with an
increase in $Y$ of 0.04, while the blue loops become longer.
The ZAHB luminosity increases of about 0.2 mag (as already noted by Sweigart
1997) and the asymptotic giant branch (AGB) reproduces, at higher
luminosity for a given \Teff, the one with the lower $Y$ (not shown).

The slower AGB phase begins with a clump due to the ignition of helium in a
shell after the end of central burning. For $Y$ = 0.245 the minimum mass to
form the helium shell on the AGB is 0.57 \Msun; larger masses form the clump
more or less at the same position in the CM diagram (\Mv\ $\sim$ -0.3 --
-0.4 mag), while lower masses form the helium shell at \BMV\ $\leq$\ 0 mag
and reach the AGB at a luminosity the higher, the lower the mass. Masses $<$
0.52 \Msun\ do not develop an AGB phase.

The shell formation phase lasts about 1.5 -- 2.5 $10^6$ yr (the lower the
mass, the shorter the phase) and another 8 -- 9 $10^6$ yr are spent by the
star along the AG branch until the first helium shell flash. The lumimosity
at which the first flash develops depends slightly on the star mass: it
varies from \Mv\ $\sim$ -2.2 mag (log($L/L_\odot$) $\sim$ 3.05) to -2.6 mag
(log($L/L_\odot$) $\sim$ 3.3), for an evolving mass of 0.52 and 0.70 \Msun,
respectively.

When compared with the case with $Y$ = 0.245, horizontal branches with $Y$ =
0.345 show noticeable differences. The luminosity level at the RR Lyrae gap
turns out to be about 0.5 mag more luminous. The tracks of the larger masses
show the well known long blue loops (Sweigart and Gross 1976). The tracks
lying in the RR Lyrae domain when $Y$ = 0.245 move to the blue of the
variable gap when $Y$ increases to 0.345 (see the model with $M$ = 0.64
\Msun).

As for the blue HB, here the effect of the helium content increase is more
limited, given the lesser efficiency of the hydrogen shell. The higher
molecular weight in the envelope produces higher effective temperatures for a
given mass, and so a lower \Mv\ due to the larger BCs: so we have that
hot structures with larger $Y$ turn out to be less luminous in the
visual than their lower helium counterparts (Fig. \ref{H2445F1}).

\begin{figure}
\hspace{1cm}
\psfig{figure=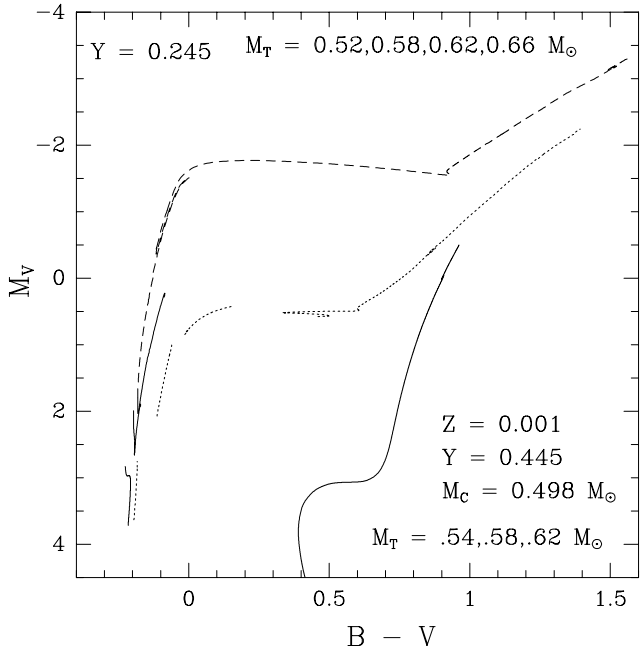,height=9cm}
\caption{Evolutionary tracks for HB models with envelope $Y$ = 0.445; for $M$
= 0.62 \Msun\ the track is extended to the AGB (dashed line: fast evolution).
As reference, HB tracks for $Y$ = 0.245 are shown (dotted line), together
with a red giant track for $M$ = 0.9 \Msun\ (solid line).}
\label{H2445F4}
\end{figure}

The minimum mass for which the shell forms on the AGB is 0.60 \Msun, at \Mv\
$\simeq$\ -0.6 mag (for comparison, $M$ = 0.7 \Msun\ forms the helium shell
at \Mv\ $\simeq$\ -0.9 mag). The 0.58 \Msun\ model represents a limiting
case and forms the helium shell, with some oscillations, at the base of the
AGB. Stars with $M$ $\leq$\ 0.53 \Msun\ do not reach the AGB phase. The
total time spent on the AGB is always less than 10 $10^6$ yr. On the whole,
the AGB with $Y$ = 0.345 is bluer of $\leq$\ 0.04 mag than the locus with $Y$
= 0.245.

We consider an increase in $Y$ of 0.2 only for structures on the blue side of
the RR Lyrae gap (Fig. \ref{H2445F4}). The helium shell forms always at
high temperatures and the AGB reduces to a short and luminous phase, well
separated from the red giant branch. At the extreme blue edge, a mass of 0.51
\Msun\ (envelope mass = 0.012 \Msun) has a ZAHB magnitude \Mv\ = 4.32, to be
compared with a ZAHB magnitude \Mv\ = 4.40 for 0.501 \Msun\ with envelope
$Y$ = 0.245 (envelope mass = 0.003 \Msun; see Fig. \ref{H2445F7}).

\section{Comparison with HB populations}
\subsection{The case of M3 and M13}

The clusters M3 and M13 offer an interesting possibility of testing the
relevance of these departures from standard HB evolution.
These clusters have quite different HB populations, notwithstanding
the great similarity in the main physical and chemical parameters (see f.e.,
Catelan \& de Freitas Pachego 1995, Ferraro et al. 1997a). Apparently, they
differ only in the ellipticity (M13 being more elliptical) and in the
presence of more rapid rotators in M13 than in M3 (Peterson et al. 1995, Behr
et al. 2000a). In brief, M3 has a HB well populated from the region redder
than the RR Lyrae gap to the blue of the gap itself, while M13 is populated
only on the blue side of the gap, with a substantial portion of HB members
populating the B subdwarf region.

\begin{figure}
\hspace{1cm}
\psfig{figure=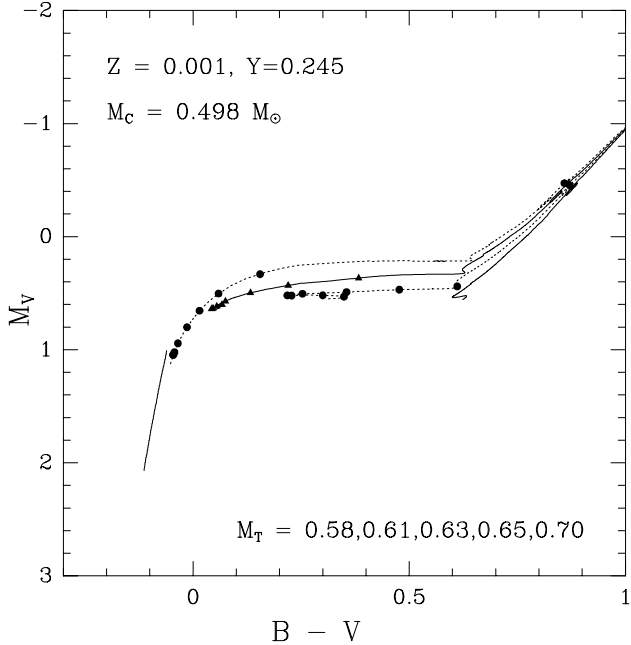,height=9cm}
\caption{HB evolutionary tracks illustrating the different mass distributions
in M3 and M13 once compared with the CM diagrams of the two clusters. Dots
and triangles indicate time steps of $10^7$.}
\label{H2445F5}
\end{figure}

M3 and M13 appear to have the same heavy element content (f.e., Kraft et al.
1992, Carretta \& Gratton 1997). Assuming such a content to be $Z$ = 0.001
(Carretta \& Gratton 1997), if also the helium content $Y$ is the same in
the two clusters, we can estimate that the bulk of HB objects in M3 has total
mass between 0.68 and 0.62 \Msun\ (see the CM diagrams by Buonanno et al.
1994, Ferraro et al. 1997a,b, and the models by Mazzitelli et al. 1995, Caloi
et al. 1997, but any other model set would do). In the same conditions, for
M13 we expect HB objects from 0.61 to 0.5 \Msun\ (see the CM diagram by
Paltrinieri et al. 1998; Fig. \ref{H2445F5}).

The origin of such a difference is uncertain.
The age difference between the two clusters is generally estimated not larger
than 1 Gyr, but values up to 3 Gyr cannot be excluded (VandenBerg et al 1990,
Catelan \& de Freitas Pachego 1995, Stetson et al. 1996, Ferraro et al.
1997a, Johnson \& Bolte 1998): Lee et al. (1994) in fact argued that age is
the most important second parameter among globular clusters.

On the other hand, the absence of RR Lyrae variables and the very low mass of
the extreme HB members in M13 (about 0.5 \Msun), suggested by their \Mv\
larger than the turn-off magnitude, would require an age difference between
M3 and M13 of various Gyr (see the discussion in Catelan \& de Freitas
Pachego 1995). Alternatively, mass loss along the giant branch would have to
differ strongly in the two cases.

In Sweigart's investigation on deep mixing (1997), to each value of helium
enhancement correspond an increase in the GB tip luminosity and an increase
in mass loss, while the size of the hydrogen exhausted core remains almost
unchanged. It is possible therefore to associate to any position
on the standard HB the shift that would result if a certain amount
of deep mixing had taken place. Given two globular clusters where deep mixing
has occurred at different degrees, it is so possible to predict -
at least semi-quantitatively - the difference between the two HB
distributions.

\begin{figure}
\hspace{1cm}
\psfig{figure=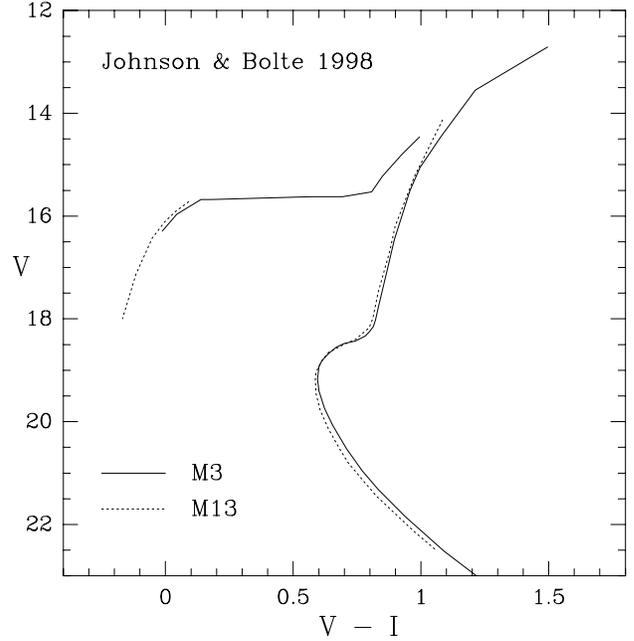,height=9cm}
\caption{The observed loci of M3 and M13 in the $V, V - I$ plane from
Johnson \& Bolte 1998, after a shift in M13 data of + 0.6 mag in $V$.}
\label{H2445F6}
\end{figure}

To this respect, the relative positions of the horizontal branches in the two
clusters are crucial. The main evolutionary loci of the two clusters have
been compared by various authors (f.e., Catelan \& de Freitas Pachego 1995,
Ferraro et al. 1997a, Johnson \& Bolte 1998). In Fig. \ref{H2445F6} we
report
the data from Johnson \& Bolte (1998) paper, to show that the two CM
diagrams $V, V - I$ can be safely superposed (keeping in mind the cautions
exemplified in Fig. 7 of the quoted paper).

Similarly, Ferraro et al. (1997a) find that their $Hubble$ $Space$
$Telescope$ CM diagrams $V, U - V$ for M3 and M13 overlap nicely, once the HB
mean loci are aligned relying on the (small) region where the HB is well
populated in both clusters. Only the reddest HB members in M13 appear more
luminous than stars in M3 at the same colour, and are interpreted by Catelan
and de Freitas Pachego (1995) and Ferraro et al. (1997a) as being more
evolved than corresponding stars in M3.

Estimates of the luminosity level of turn-off and HB for M3 and M13 are given
by Buonanno et al. (1994), Ferraro et al. (1997b), Paltrinieri et al.
(1998). Also these data do not show any noticeable difference in the
relative positions of the evolutionary loci in the two clusters.

Therefore, the difference (if any) in luminosity between the HBs in M3 and
M13 can be safely assumed to be lower than 0.2 mag. This means that the
helium enhancement for the coolest M13 HB members cannot exceed $\delta Y$ =
0.04 (Sweigart 1997 and Section 2 of this paper). The extra mass loss in the
GB phase due to this increase in $Y$ in the red giant envelope is of about
0.012 \Msun\ (from Fig. 3 in Sweigart 1997, assuming $\eta$ = 0.4 and keeping
in mind the value for the mixing parameter quoted by the author to give an
increase in $Y$ of 0.04). A star on the red side of the HB in M3 has a mass
of about 0.68 \Msun\ (for $Z$ about 0.001, see Fig. \ref{H2445F5})
and, for the same metallicity, a star at $B - V$ = 0.1 in M13 has a mass of
about 0.61 - 0.62 with the same $Y$ (0.245) or of about 0.64 - 0.65 for $Y$ =
0.285. An extra mass loss of 0.012 \Msun\ appears not sufficient to allow the
disappearance of the RR Lyrae variables (see Fig. \ref{H2445F1} and Fig.
\ref{H2445F5}).

\begin{figure}
\hspace{1cm}
\psfig{figure=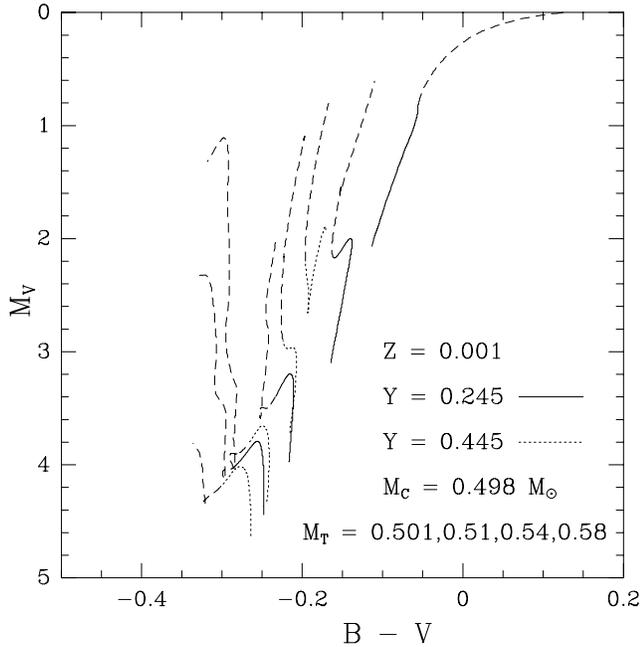,height=9cm}
\caption{Evolutionary tracks for HB models with envelope $Y$ = 0.445 (dotted
line) and 0.245 (solid line); dashed lines indicate fast phases. With
increasing mass the difference in luminosity and colour between models with
the same mass increases substantially.}
\label{H2445F7}
\end{figure}

Higher helium mixing and larger mass losses may be present in very blue HB
members ($B - V$ $\leq$ -0.1). In Fig. \ref{H2445F7} the effect of an
increase $\delta Y$ = 0.2
in the envelope of extremely blue HB objects is shown. In order to
reach an \Mv\ $\simeq$ 4.5 mag (as in M13 and NGC 6752, where the HB
reaches visual magnitudes higher than the turn-off), a total mass as low
as 0.501 \Msun\ is needed for $Y$ = 0.245, while for $Y$ = 0.445 a mass
$\leq$ 0.51 \Msun\ is required. So even for extensive mixing one finds that
a substantial mass loss is necessary to attain the high magnitudes of extreme
HB objects.

Such mass loss turns out larger than expected from the increase in the GB tip
luminosity. From Fig. 3 in Sweigart's paper, one gets an extra mass loss of
$\sim$ 0.08 \Msun\ when $\Delta X_{\rm mix}$ = 0.20 ($\delta Y$ $\sim$
0.2); so the minimum mass in
M3 ($\sim$ 0.62 \Msun) would become 0.54 \Msun, definitively larger than the
required 0.51 \Msun\ (see Fig. \ref{H2445F7}).

On the whole, an increase in $Y_{\rm env}$ is specially efficient in shifting
the position on the HB for the mass range of the RR Lyrae variables and for
slightly lower masses (see Fig. \ref{H2445F1}), but at the price of a
noticeable increase in luminosity. Anyway, it is important to keep in mind
that this discussion is based on the assumption that mass loss follows
strictly Reimer's formulation all along the giant branch. Should mass loss
increase more rapidly with luminosity, the conclusions could be different.

\subsection{The case of M13 and NGC 7006}

Kraft et al. (1998) apply the Sweigart (1997) models to estimate the required
amount of helium enhancement in M13 with respect to NGC 7006, in order to
achieve the observed differences in HB morphology. The required helium
abundances are 0.35 and 0.245 for M13 and NGC 7006, respectively. Fig.
\ref{H2445F1} shows that such abundances require for $\Delta_{\rm
ZAHB}^{TO}$ in M13 a value 0.5 mag larger than in NGC 7006, if the two
clusters have similar ages (similar turn-off levels). Instead, one finds 3.50
$\pm$ 0.10 mag for both M13 (see Fig. \ref{H2445F6} ) and NGC 7006
(Buonanno et al. 1991). So it appears difficult to accomodate a difference in
HB level of 0.5 mag, without considering an age difference of more than 5 Gyr
in the sense of a younger M13 (since its turn-off would appear more luminous
than in NGC 7006).

\subsection{The metal poor clusters}

This difficulty presents itself also in the case of M15,
the metal poor globular cluster with many red giants interested by the
signature of deep mixing (Cohen 1979, Sneden et al. 1991, Sneden et al.
1997). Sweigart (1997) observes that an increase in $Y_{\rm env}$ of 0.04
would give rise to a luminosity difference between the HBs in M3 and M15
which would explain the period-shift effect (Sandage 1982). The luminosity
increase would be of about 0.2 mag.

The comparison with CM diagrams of other very metal poor clusters can give
some clues as to the reality of such an increase in $Y_{\rm env}$. Walker
(1994) compared carefully the CM diagrams of M15 and of M68, showing that the
two diagrams overlap nicely, and that, in particular, the difference in
magnitude between HB and TO $\Delta_{\rm ZAHB}^{TO}$ is the same in the two
clusters (3.41 $\pm$ 0.05 mag). Similarly, Rey et al. (1998) found for
$\Delta_{\rm HB}^{TO}$ in M53 the value of 3.42 $\pm$ 0.05 mag. At the same
time, the difference in age among the three clusters appear definitively
smaller than 2 Gyr.

This indicates that the luminosity of M15 HB is the {\it same} as found in
M68 and M53.
Nothing is known as to peculiar or not peculiar abundances in M68 and
M53 giants, but they have no blue tail in their HB
distributions, and so no hint of extra mass loss on the giant branch or
extra helium in their HB structures. Besides, Brocato et al. (1994)
and Walker (1994) find that the pulsational properties of M68 RR Lyrae
variables are analogous to those in M15, suggesting similar luminosities.

It seems therefore that the population on M15 HB in the region of the RR
Lyrae stars resembles quite strictly the population in apparently normal
clusters such as M68 and M53. The amount of extra helium $\delta Y$ allowed
in M15 structures around the RR Lyrae gap is surely less than 0.04, which
would imply a luminosity increase of 0.2 mag with respect to the HBs in M68
and M53. Stars with helium enhanced may be found in bluer positions than the
gap surroundings.

Summing up, there are no indications of a substantial increase in the helium
content in most of the HB stars in clusters whose giants show evidence of
mixing of envelope material into helium rich regions, such as those where Na
and Al enrichment take place.

\section{Deep mixing: first or asymptotic giant branch?}

The situation with the variety of relative abundances of CNO, Na, Mg, Al,
{\it etc} in globular clusters has reached a high degree of complexity
(see f.e., the review by Kraft 1994). There is evidence for primordial
fluctuations (f.e., CN versus CH anticorrelation in main sequence and
subgiant stars in NGC 6752, 47 Tuc and M92: Suntzeff 1989, Bell et al. 1983,
Briley et al. 1991, King et al. 1998) as well as for evolution induced
changes in surface abundances of various atomic species (CNO, Na, Al, Mg and
their isotopes: quoted paper by Kraft). A large dispersion in O abundances
is even observed in a young cluster in the Small Magellanic Cloud, NGC 330
(Hill 1999).

We would like to consider only one aspect of this vast problem: the
evolutionary status of the most affected objects in one of the best studied
clusters, M13. The most luminous giants show strong O deficiency and
Na overabundance. It is commonly assumed that these giants
belong to the first giant branch, on their way to the helium flash.

We suggest on the contrary that a substantial fraction of the objects
observed near the giant tip are asymptotic branch stars. This statement is
mainly based on the fact that the majority of the luminous giants in M13 are
variables of the semiregular (SR) type with period longer than 30 days,
typical of the last phases of the AGB in intermediate and very metal poor
globular clusters (f.e., Feast 1999). While Mira variables lie above the red
giant tip as defined by the developing of the helium flash, and so we can be
sure that they belong to the AGB phase, red SR variables are not always above
the RG tip. Nevertheless, it seems likely that they are AGB stars, since: i)
there is evidence that in the metal rich globular clusters they evolve into
the Mira phase; ii) if we consider the SR variables for which periods can be
obtained with some confidence, it appears that the light variations are due
primarily to pulsation, as in the case of Miras (Feast 1999); iii) Whitelock
(1986) has shown that in globular clusters the luminosities of the SRs
increase with the period and that the sequence defined by these stars
terminates at the position of the Miras in the cluster (see also Feast 1999).

In Table 1 by Pilachowski et al. (1996), seven
out of the ten most luminous giants in M13 turn out to be variable or
probably variable: L324 = V11 in Sawyer Hogg's catalogue (SH 1973; light
curve in Osborn and Fuenmayor 1977), L598 (=B140) probably variable according
to Kadla et al. (1976), L194 (= Arp II-90) definitively variable according to
Welty (1985), L973 (= Arp I-48) light curve in Osborn and Fuenmayor (1977),
L835 = V15 in SH (light curve in Osborn and Fuenmayor), L954 (= Arp IV-25)
definitively variable according to Welty (1985),
L70 (= Arp II-67) probably variable according to Welty. L414 (= Arp III-56)
has been suspected, but not confirmed, of being variable (Russev 1973, Osborn
and Fuenmayor 1977).

Five out of the seven tip variables are super-oxygen poor ([O/Fe] $<$ -0.4)
and one is oxygen-poor ([O/Fe] = -0.26) according to Kraft et al. (1993,
Table 6). From Table 1 in Pilachowski et al. (1996) it appears that the
[Na/Fe] ratios near the red giant branch tip are exclusively high (with the
exception of L598), while on the lower giant branch there is a wider range
in Na abundances. We suggets therefore that, whatever the phenomenon
involved, it is specially active during the final evolution along the
asymptotic giant branch, and in particular in the evolution of stars with
small envelopes (as supposedly are the ones found on the AGB of M13).

The variability of Miras has been associated with the helium shell flashes
phase (Feast 1989). Since SR variables appear related to Miras, as mentioned
before, also they are likely in the terminal stage of the AGB evolution.
We computed a series of 8 flashes for a 0.65 \Msun: each flash cycle lasted
about 2 -- 3 $10^5$ yr and during it the hydrogen exhausted core
increased of about 0.009 \Msun\ and the \Mv\ at maximum of about
0.09 mag. These results have to be taken as indicative, given the well known
extremely strong dependence of thermal pulse development on stellar
structure details.

As mentioned before (Sec. 2), the flashes begin at increasing
luminosty for increasing evolving mass (from log(L/\Lsun) $\sim$ 3.05 --
\Mv\ $\sim$ -2.2 -- to $\sim$ 3.3 -- \Mv\ $\sim$ -2.6 -- for masses from 0.52
to 0.70 \Msun). These luminosities are in any case very close to the the
first GB tip: log(L/\Lsun) $\sim$ 3.38 (\Mv\ $\sim$ -2.8) for \Z\ = 0.001
(f.e., Salaris \& Cassisi 1997). If the luminosity of the AGB termination is
close, as generally accepted (Renzini 1977), to the one of the first GB,
only a few flashes will take place, for a total duration of the order of
$10^6$ yr or less. For comparison, a first GB member moves from
log(L/\Lsun) = 3.05 to the GB tip in about 3 $10^6$ yr, and from 3.3, in 6
$10^5$ yr.

So the expected evolutionary rates in luminosity along the two giant branches
are similar close to the tip. Actually, since the evolution in luminosity in
the AGB pulsing phase is roughly constant (see also Gingold 1974), it may
become slower than along the first branch.

In a cluster like M13, whose HB is populated by stars with mass $\leq$ 0.62
\Msun\ (see before), most stars will begin their pulsing phase at \Mv\
$\sim$ -2.2 -- -2.4, with a hydrogen exhausted core of about 0.52 \Msun.
An HB mass of 0.58 \Msun, increasing the core mass of 0.009 \Msun\ per flash,
will perform about 6 flashes in absence of mass loss, reaching the luminosity
of the first GB tip (\Mv\ $\sim$ -2.8).

In a cluster like M3, where the most common HB mass is about 0.65 \Msun\ (the
mass in the RR Lyrae gap), the flash phase begins at \Mv\ $\sim$ -2.6, about
0.2 mag below the theoretical tip and about where we expect to
observe the most luminous cluster giants (Frogel 1983, Salaris \& Cassisi
1997, 1998). Without mass loss, the star could perform about 14 pulses, with
an increase in luminosity of about 1.3 mag, well above the first GB tip. This
point hase been discussed by Renzini (1977), who reached the conclusion that
during the AGB the evolving mass loses at least 0.1 \Msun. Just for the sake
of the argument, we notice that the most luminous giant in M3 is a SR
variable (V95 in SH = vZ318: $V$ = 12.36 mag, $(B - V)$ = 1.78 mag, period =
103 d, SH and Welty 1985); the second one in luminosity is again a variable
(vZ1397 = SK26, $V$ = 12.65, $B - V$ = 1.56 mag, period (uncertain) = 60 d,
Welty 1985). At about this magnitude, a few more giants are found, so that
V95 results isolated 0.3 mag above the bulk of the GB members.

In Sec 3.1 we saw that M3 and M13 CM diagrams can be overlapped with a shift
in $V$ of about 0.6 mag, M3 being the farther away. The tip of M13 is at $V$
$\sim$ 12 mag (Pilachowski et al. 1996), so that we expect the tip of M3 at
$V$ = 12.6 mag, as actually observed, exception made for V95: this may be
taken as a consequence of the larger -- on the average -- AGB masses in M3
with respect to M13 (in the limits of small number statistics!).

Some advantages can be found if the mixing takes place in AGB objects. At
luminosities close to the first giant branch tip the core mass $M_{\rm c}$
would be $\sim$ 0.5 \Msun, so that an evolving giant of about 0.65 \Msun
would have an envelope mass of about 0.15 \Msun, to be mixed with the
products of (high temperature) hydrogen shell burning. In the case of an AGB
object of the same mass, undergoing helium shell flashes, the envelope mass
would be smaller, starting from about 0.13 \Msun\ (in the hypothesis of no
mass loss during AGB evolution before the flashes) and rapidly decreasing
both for mass loss and for the increase of the hydrogen exhausted core. The
abundance changes would be faster and would require the consumption of less
hydrogen.

In a cluster like M13 the evolving AGB masses will be lower than 0.65 \Msun,
so that the envelope masses to be mixed will be definitively smaller than in
the first GB case, with an evident advantage for the reshuffling of nuclear
species. Just as a curiosity: there are indications that some Type II
Cepheids exhibit abundance peculiarities, among which possibly helium
enhancement, as summarized by Gingold (1985) some years ago. It may be worth
while to reconsider the objects in question with modern equipment.

For what concerns the nucleosynthesis beyond the standard CNO cycle,
the conditions in the hydrogen shell in AGB structures are not very different
from the ones encountered in first GB stars. Peak temperatures are of about
55 - 58 $10^6$ K before the first helium shell flash and slightly higher (58
- 60) $10^6$ K during the eighth flash (0.65 \Msun, see before). The peaks
of CNO energy generation
during the flash cycles last about 2 $10^5$ yr, sufficient to obtain large
sodium and aluminum enhancements in surface abundance (Langer \& Hoffman
1995). A difficulty is given by the estimate by Langer et al. (1997) of a
temperature of 70 $10^6$ K necessary to explain surface abundances of Mg
isotopes as observed by Shetrone (1995, 1996) in the most luminous M13
giants. As stressed by Langer et al. (1997), the estimate of 70 $10^6$ K
depends very strongly on these results, for which a confirmation would be
welcome for safety. In any case, such a difficulty is present for both the
RGB and the AGB hypotheses.

\section{An old conjecture on the origin of extended blue tails}

The evolution along the GB suffers a strong acceleration
when approaching the helium flash phase. The rate of mass loss, estimated
through Reimer's formula, reaches values of about 5 $10^{-8}$ \Msun ${\rm
yr^{-1}}$ ($\eta$ $\sim$ 0.4). In the last million year before the flash,
models of 0.8 -- 0.9 \Msun\ ($Z$ $\leq$ 0.001) have an increase in core mass
of $\sim$ 0.027 \Msun\ and a decrease in total mass of 0.05 \Msun. This
means
that a variation of the timing of the helium flash of half a million
year, that is, about 2\% of the time to climb the giant branch from
the RG bump luminosity to the tip (30 -- 40 Myr), is sufficient to move a
structure from the red HB section to the RR Lyrae region (for $Z$ $\sim$
0.001).

\begin{figure}
\hspace{1cm}
\psfig{figure=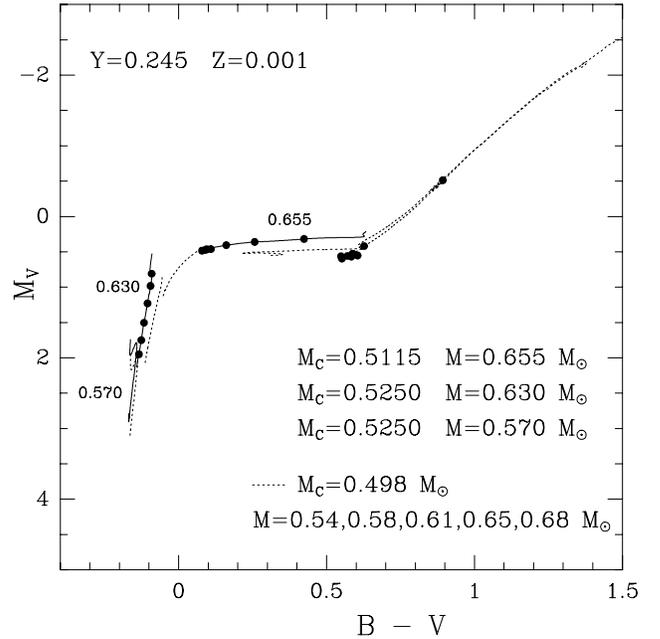,height=9cm}
\caption{Evolutionary tracks for HB models with standard envelope $Y$ =
0.245 and $M_{\rm c}$ = 0.498 \Msun\ (dotted line) are compared with tracks
supposedly resulting from an evolution beyond the standard GB tip of 5
$10^5$ yr for 0.655 \Msun, and $10^6$ yr for 0.63 and 0.57 \Msun. Dots
indicate time intervals of $10^7$ yr.}
\label{H2445F8}
\end{figure}

This is a well known fact, and has been exploited to invoke rotation as a
possible "second parameter" in shaping HB morphology, since the increased
pressure delays core heating (Mengel \& Gross 1976, Renzini 1977).

The question of rotation and its relevance to HB morphology has been recently
investigated (Peterson et al. 1995, Behr et al. 2000a,b).
An overall correspondence is found between {\it v} sin{\it i} and
position on the HB, in a sense perhaps contrary to what expected. So stellar
rotation rate is clearly related to the position on the HB, but in a way not
yet understood.

Independently from the causes, let us consider the consequences that a {\it
small} delay in the helium flash timing can have on the HB morphology of a
cluster like M3. As mentioned before, the maximum and minimum HB masses for
the bulk of the HB population in this cluster can be assumed to be 0.68 and
0.62 \Msun (\Z = 0.001, $M_{\rm c}$ = 0.498 \Msun). The maximum mass,
after an extra 0.5 $10^6$ yr spent near the GB tip, reduces to 0.655 \Msun,
while $M_{\rm c}$ increases to 0.5115 \Msun; similarly, after $10^6$ yr one
has a stellar structure of total mass 0.63 \Msun\ and $M_{\rm c}$ = 0.525.

In Fig. \ref{H2445F8} the evolutionary tracks of these structures are
shown, together with the track derived from 0.62 \Msun\ after $10^6$ yr of
extra GB evolution ($M$ = 0.57 \Msun, $M_{\rm c}$ = 0.525 \Msun), and the
tracks (until $Y_{\rm c}$ = 0.1) of "standard'' HB stars ($M_{\rm c}$ = 0.498
\Msun). One finds that an extra GB evolution of 0.5 $10^6$ yr is sufficient
to transform the HB of M3 into the one observed in M2 (Lee \& Carney
1999a,b), and that an extra of $10^6$ yr transforms the HB morphology of M3
into the one of M13, exception made for the bluest members.

The extremely blue HB stars would require longer extra time, but never
larger than 1.5 $10^6$ yr, since such an extra evolution leaves a structure
with about 0.006 \Msun\ in the envelope. In these conditions, the subsequent
evolution is the one of a "hot flasher" or flasher along the white dwarf
cooling sequence, as discovered by Castellani \& Castellani (1993) and
investigated also by D'Cruz et al. (1996). Such structures are found at the
very blue end of foreseable HB distributions (see the quoted papers).

\section{Conclusions}

The peculiar chemical abundances in many high luminosity red giants in
globular clusters apparently require an increase in their envelope helium
content. It was not possible to find a confirmation of such an increase from
the position of the ensuing horizontal branches: the shift from the red to
the blue of the RR Lyrae variables gap requires a substantial increase in
luminosity, inconsistent with observations.

The possibility is examined that the most peculiar giants belong to the
second, or asymptotic, giant branch, and precisely to the phase of the helium
shell flashes. Some support for the hypothesis is found in M13, where the
majority of the peculiar stars are variables of the long period, semiregular
type.

An estimate is given of the consequences of a delay of the helium flash, on
the basis of recent RG models: a difference in the flash timing between 0.5
and 1.5 $10^6$ yr can help to explain at least part of the shift from the M3
HB morphology to the one of M2, and the appearance of the extreme blue tails
(M13, NGC 6752). The many features which accompany the appearance of gaps
and blue tails in HBs remain in any case unexplained.


\end{document}